\documentclass[sigconf]{acmart}
\AtBeginDocument{%
  \providecommand\BibTeX{{%
    \normalfont B\kern-0.5em{\scshape i\kern-0.25em b}\kern-0.8em\TeX}}}

\copyrightyear{2022}
\acmYear{2022}
\setcopyright{acmlicensed}\acmConference[CHI '22]{CHI Conference on Human Factors in Computing Systems}{April 29-May 5, 2022}{New Orleans, LA, USA}
\acmBooktitle{CHI Conference on Human Factors in Computing Systems (CHI '22), April 29-May 5, 2022, New Orleans, LA, USA}
\acmPrice{15.00}
\acmDOI{10.1145/3491102.3517675}
\acmISBN{978-1-4503-9157-3/22/04}

\newcommand{\change}[1]{#1}
\newcommand{\minorchange}[1]{#1}
\newcommand{\userx}{$u_{x}$}
\newcommand{\userb}{$u_{b}$}
\newcommand{\usero}{$u_{o}$}
\newcommand{\tweetx}{$t_{x}$}
\newcommand{\tweeto}{$t_{o}$}

\begin{document}

\title{Understanding Privacy Switching Behaviour on Twitter}


\author{Dilara Kek\"{u}ll\"{u}o\u{g}lu}
\affiliation{\institution{University of Edinburgh} \country{Edinburgh, UK}}
\email{d.kekulluoglu@ed.ac.uk}

\author{Kami Vaniea}
\affiliation{\institution{University of Edinburgh} \country{Edinburgh, UK}}
\email{kvaniea@inf.ed.ac.uk}

\author{Walid Magdy}
\affiliation{\institution{University of Edinburgh} \country{Edinburgh, UK}}
\email{wmagdy@inf.ed.ac.uk}

\renewcommand{\shortauthors}{Kek\"{u}ll\"{u}o\u{g}lu et al.}

\begin{abstract}

Changing a Twitter account’s privacy setting between public and protected changes the visibility of past tweets. By inspecting the privacy setting of more than 100K Twitter users over 3 months, we noticed that over 40\% of those users changed their privacy setting at least once with around 16\% changing it over 5 times. This observation motivated us to explore the reasons why people switch their privacy settings. We studied these switching phenomena quantitatively by comparing the tweeting behaviour of users when public vs protected, and qualitatively using two follow-up surveys (n=100, \change{n=}324) to understand potential reasoning behind the observed behaviours. Our quantitative analysis shows that users who switch privacy settings mention \change{others} and share hashtags more when their setting is public. Our surveys highlighted that users turn protected to share personal content and regulate boundaries while they turn public to interact with others in ways the protected setting prevents.

\end{abstract}



\maketitle

\section{Introduction}

Interpersonal boundaries are continuously shifting concepts for people as they move between situations, contexts, and as the world they exist in changes. Humans are social creatures and are well accustomed to this continuous shifting of how they present themselves and how they manage their boundaries around personal privacy, self presentation, and access to self. The introduction of social media platforms though can lead to challenges in boundary management where many of the natural high-granularity actions taken in the physical world must be translated into the more rigid privacy setting options used by platforms~\cite{ellison2011negotiating}. \change{While many people either do not change the default settings provided by the platforms~\cite{fiesler2017or} or set them only once during account creation~\cite{strater2008strategies}, other} people have found a myriad of ways to use \change{social media platforms'} rigid setting structure\change{s} to mimic the fine gradations of inter-personal boundaries, often in unexpected and non-obvious ways. In this work, we look at Twitter's tweet visibility setting feature which is particularly ridged with only two options, public and protected, which apply to the user's entire tweet stream, including historical tweets. Yet even in this highly binary and low fidelity situation we see a range of how people use the system to create a rich boundary management to suit their needs. 

Twitter is a particularly interesting platform to study in regards to privacy management because of the limited control options and also because Twitter is quite blunt in its help center about the impact of switching from protected to public: ``Unprotecting your Tweets will cause any previously protected Tweets to be made public''~\cite{twitter-public-protected}. The implication is that Twitter accounts are meant to be public or protected and are generally assumed to stay in one of those states with changes between being rare. 
Existing work on privacy settings on Twitter mostly focus on the differences between users who are protected and public~\cite{choi2015self,liang2017privacy}. These studies check the settings of user accounts only once to decide \change{if it is} public or protected. However this assumption may be wrong and it may be the case that users are changing their account visibility often enough to count as having a more complicated visibility status.

Studies in other platforms, like Facebook, have found that even when privacy controls do not well match the effects users are trying to create, people are quite good at using the technology in unexpected ways to create the effects they want. A good example is work on how teenagers manage context in social networks by doing things like deactivating their accounts when they are not logged in~\cite{marwick2014networked}. While not immediately intuitive, deactivating an account prevents others from interacting with it, enabling the user to only allow others to interact with the account when they themselves are logged in. The point is that privacy permission management is not necessarily a single set-it-and-forget-it setting. People use these settings in boundary management, which is a continuously changing state between people. 

While users may need to manage interpersonal relationships with high granularity, the main control over privacy on online social networks is still the visibility of the account content. 
On Twitter, that means the visibility of the account and its tweets. Twitter users have a relatively small set of options to protect their tweets and protected user accounts also face restrictions on the types of interactions they can have. For example, if a protected user mentions a non-follower, that person cannot see the mention and therefore cannot respond to it. Other than changing tweet visibility, users can also delete their tweets, block other users to prevent them from interacting, and deactivate their accounts. 

It is also an open question what effect users are attempting to create when they change visibility settings. The most obvious assumption is that they are trying to hide their tweets from public consumption. But Twitter also has other issues, such as harassment~\cite{jhaver2018online}, that may cause users to switch to protected~\cite{veletsianos2018women}. This observation opens an interesting question about how users behave when they are public vs when they are protected. A user simply trying to avoid harassment may actually not change behaviour when they are protected, because it is themselves they are trying to protect, not the tweets. Users that are trying to protect their tweets may also have specific types of tweets that they are more inclined to post while protected \change{as opposed to when} public.

In this work, we focus on users who change their privacy settings in order to understand the behaviour and motivations behind the changes, as well as the differences in self-disclosure between when they are public versus protected. More precisely, we investigate the following research questions:
\begin{description}
    \item[RQ1] How frequently do Twitter users change their tweet visibility settings?
    \item[RQ2] Do users employ different posting strategies when they are public vs protected?
    \item[RQ3] Why do users change their visibility settings and how do the reasons differ between changing to protected versus public?
    \item[RQ4] What other strategies do switching users employ to control their audience and interactions?
    
\end{description}

To answer these questions, we curated a set of 107K users\change{, whose accounts were protected at the start of the data collection,} and analyzed their account visibility changes for three months. Of these, 45K changed their visibility to public at least once allowing us to collect their tweets to determine if their tweeting behaviour changes when they were protected compared to when they were public. We also conducted two user surveys to get insights into why users decide to change their tweet visibility back and forth.

Our findings show that a large portion of protected users do change their privacy setting to public, around 40\% of the protected accounts we inspected changed to public at least once within three months. A quarter of those \change{accounts} changed more than ten times within this period. 
\change{Analysis of tweet data shows that users' accounts have less tweets during times when they are protected. They also mention other users less compared to when they are public.}
Our survey results show that users mostly change their tweets to protected to regulate their boundaries but change to public to overcome the platform's technical constraints. 
\change{Our findings provide a unique perspective into the usage of privacy settings in a platform where options are binary and in an account-level, where trade-off between privacy and functionality must be calculated with the historical posts in mind. We also provide design implications built upon our findings and prior literature to help users and platforms minimize potential privacy leaks.}

\section{Related Work}

\minorchange{People use online social network platforms in a range of ways including} connecting with others, gaining social capital, and learning about events, all while also wanting to protect their privacy~\cite{ellison2011negotiating,vitak2015balancing}. People also use social media to seek support~\cite{dym2018vulnerable,andalibi2016understanding}, build reputation~\cite{syn2015social}, and for professional contexts~\cite{mahrt2014twitter,marwick2011tweet}. People can choose to only connect with a specific group of people, or multiple social circles. This usage can cause context collapse where these social circles, such as friends, family, and colleagues, are all on the same social network~\cite{nissenbaum2009privacy,marwick2011tweet,vitak2012impact} making it hard for users to control the flow of information. 

Altman~\cite{altman1976conceptual} describes privacy as a dynamic process where people select what part of themselves they want to show, when they want to show it, and who they want to show it. It is not trivial to regulate this on social media where posts are usually permanent~\cite{madejski2011failure}. Especially on Twitter where privacy settings are not granular and once changed they apply to all of the posts shared in the past. Once shared, the information becomes co-owned with the selected audience~\cite{petronio2015communication}. Hence, it is critical to select what information to impart and the audience that will see it on social media.

\subsection{Usage of Privacy Settings}

\change{Social network platforms provide privacy settings for users to adjust a range of configurations including setting the audience of their posts, regulating interactions, changing the visibility of shared information, and so on. However, users keep the defaults provided by the system~\cite{fiesler2017or,liu2014tweets,gross2005information} for various reasons such as difficulty of finding and using the settings~\cite{van2015social,sleeper2013post}, uncertainty about their preferences~\cite{acquisti2020secrets}, or not trusting the system to keep the configured settings~\cite{sleeper2013post}. Some users choose not to change their settings even when there are posts shown to be violating their privacy~\cite{madejski2012study,mondal2019moving} and some configure their settings only during initial setup~\cite{strater2008strategies}.}

\change{Conversely, some users} utilize privacy configurations given by platforms to protect their privacy online~\cite{marwick2014networked, ellison2011negotiating,madejski2011failure}. However, configuring privacy settings to accurately reflect the user's intended audience is not trivial~\cite{liu2011analyzing,madejski2011failure,hoyle2017viewing}. Privacy violations in social media may be caused by unclear permissions or people not knowing how the permissions work. Liu et al.~\cite{liu2011analyzing} analyzed the privacy settings set by Facebook users and compared them to the desired settings to understand the discrepancies. They found that users' expectations did not reflect the actual settings of their posts most of the time and that the size of their audience was also larger than expected.

Fiesler et al.~\cite{fiesler2017or} compares public and private Facebook posts to analyze what information is shared and who they are shared with. They collect recent posts of 1,815 Facebook users and analyze the contents as well as their privacy settings. They find that public posts and non-public posts have similar content but users with certain demographics tend to utilize the settings. Liang et al.~\cite{liang2017privacy} compares public and protected users in Twitter and find that users who use privacy settings are more likely to disclose geo-location in their tweets. 

Privacy settings provided by platforms have different granularities. For example, Facebook allows users to configure each post's audience where in Twitter the settings applied to the whole account. Hence, users might utilize them in different ways. Choi and Bazarova~\cite{choi2015self} compares users' behaviour on Facebook and Twitter. They find that while users share more intimate content on Facebook, there are no differences between protected and public Twitter users in terms of content. Similarly Liang et al.~\cite{liang2017privacy}, find that public Twitter users had more privacy concerns than the protected ones since the latter have more clear and less permeable boundaries. 

Most prior work compares public and protected user behavior using a between-subjects design. That is, they look at what users who are private and users who are public do. These studies typically find that public and protected users generally share similar things~\cite{fiesler2017or,choi2015self}. However there is an obvious gap here as it may be that the same user switching between public and protected does not share in a similar way. In this work, we use a within-subject design where we consider how behavior changes when a user changes their visibility.

\subsection{Controlling Content and Interactions}
Other than configuring the privacy settings, users utilize various methods to protect their privacy online including self-censorship~\cite{marwick2011tweet,das2013self,sleeper2013post}, limiting connections~\cite{ellison2011negotiating,johnson2012facebook}, creating multiple accounts for different purposes~\cite{vitak2015balancing,stutzman2012boundary,xiao2020random}, as well as deactivating~\cite{marwick2014networked}. Users sometimes take drastic measures such as stopping usage completely~\cite{baumer2013limiting, grandhi2019stay,lampe2013users} to protect their privacy, which can have effects on their social connections~\cite{grandhi2019stay}, and professional life~\cite{baumer2018departing}. Cutting off connection can also be harmful for vulnerable populations~\cite{dym2018vulnerable}. Lampinen et al.~\cite{lampinen2011} interviewed 27 participants about online sharing behaviours and boundary regulation tactics. They divide their identified set of strategies into \emph{preventative} and \emph{corrective}. Preventive strategies included not sharing particular contents, or sharing content only with a targeted audience. Corrective strategies included deleting an already shared post or presenting it as a joke.

Online content and interactions has a persistence over time that is not present in offline interactions. Users can choose to delete some or all of their posts for various reasons such as regret~\cite{sleeper2013read}, typos~\cite{almuhimedi2013tweets}, to prevent temporal context collapse~\cite{huang2020you}, and so on.
However, the residual activity around the deleted/hidden tweets can still be used to infer the content~\cite{mondal2016forgetting,kekulluoglu2020analysing,kekulluoglu2022from,almuhimedi2013tweets}. Almuhimedi et al.~\cite{almuhimedi2013tweets} followed 292K users for a week to analyze deleted tweets. They found that around half of the users deleted at least some tweets during the week. The content of the tweets that were deleted were not substantially different than the ones that were left on the platform. Mondal et al.~\cite{mondal2016forgetting} analyzed the accessibility of tweets longitudinally by trying to recollect tweets from older datasets that date back to six years. Unlike Almuhimedi et al.~\cite{almuhimedi2013tweets}, they also included the tweets of users who turned their accounts to protected to their analysis. They found that while most of the users did not withdraw any of their tweets, 8.3\% of them deleted some tweets, 16\% deleted their account, and 10\% protected their tweets. 
They found that they can recover the topics of interests of the withdrawn tweets by analyzing residual interactions around them which shows that even deletion might not prevent possible privacy risks.

Users also try to sanction interactions without harming their relationships to regulate \minorchange{boundaries}. Rashidi et al.~\cite{rashidi2020s}  interviewed 23 young adults to analyze the sanctions enforced by them on popular social media \minorchange{platforms}. They grouped the sanctions using three dimensions: on-site and off-site, individual and collaborative, visible and invisible sanctions. The found that people prefer using invisible sanctions like muting rather than blocking, unfriending, or deleting content.

Instead of sanctioning, users may choose to deactivate their accounts to protect their privacy, concentrate on their work~\cite{baumer2013limiting}, and to limit interactions~\cite{marwick2014networked}. Platforms provide deactivation as an alternative to deleting accounts and users are allowed to get their accounts back if they decide to do so. For example, Twitter gives a 30 day grace period during which users can reactivate their deactivated accounts. After that time the accounts will be permanently deleted and the username can be claimed by someone else.  Ng~\cite{ng2020re} describes the users who temporarily deactivated their accounts as intermittent discontinuers and found that social media fatigue was one of the reasons of the deactivation. They also found that some users change their tweets to protected instead of deactivating.

\section{Monitoring Switching Behaviour on Twitter}
\label{sec:collected}

To understand the phenomena of users switching their privacy settings on Twitter, we initially \change{curated} a large set of users who ha\change{d} protected accounts \change{at the start of the data collection} and monitored their behaviour over \change{three months} to record any possible switch to their status between protected and public. 
We analyzed these switching patterns and compared their tweets using features such as sharing of hashtags, media, links, and so on. In this section, we describe the Twitter API as well as our data collection and analysis. 

\subsection{Ethical Considerations}
We understand that even though we access the tweets of public accounts, users may have more fluid account types as shown in this study. While Twitter is quite clear about the implications of making an account public and most of the survey participants evidenced awareness that their tweets can be accessed by non-followers when they changed their tweets to public, it might not be clear to the users that researchers are also included on that set~\cite{fiesler2018participant}. To mitigate possible risks, we only report on the metadata curated from the social media data. We do not single-out any users and avoid sharing their social media content to protect their privacy. We also do not collect social media usernames or data of the survey takers. Only the researchers on this project have access to the collected raw data. We followed our University's ethics protocol in the design and running of this work, including the social media data collection and following surveys.

\subsection{Twitter Functionality}

Twitter has a binary privacy configuration where accounts are either public (default) or protected. All tweets associated with a \emph{public} account can be seen by anyone on the Internet, while all tweets associated with a \emph{protected} account can only be seen by the followers of that account. Changing an account's setting changes the visibility of all tweets associated with this account, including past tweets. Any Twitter user can follow a \emph{public} account, while following a \emph{protected} one first requires the account owner's approval. A \emph{protected} account's tweets can be liked by the followers but they cannot be retweeted or quote tweeted. Interactions of \emph{protected} accounts can only be seen by the followers. Hence, if these \emph{protected} accounts reply to a public tweet of a non-follower, that will not be visible to that non-follower. This also applies to mentioning users by using their handles (@username), quote tweeting them, as well as liking and retweeting them. Hence, \emph{protected} accounts must change to \emph{public} if they want their interactions with non-followers to be seen, but doing so also makes their full tweet history public. On the other hand, if a \emph{public} account mentions a \emph{protected} one, anyone on the Internet can see the content since the owner of the tweet is \emph{public}. We use tweet visibility and account visibility interchangeably in this paper.

\subsection{Collecting Switching Users}
As mentioned, Twitter accounts are by default public and users need to actively change their account to protected if they want to limit their tweet visibility. Since our aim is to study users who switch between public and protected, we collected users with protected accounts which indicate that they have changed their tweet visibility at least once. To obtain our sample of protected users, we first sampled the Twitter stream without any filters and collected a list of mentioned protected users from these tweets for 60 hours starting on the 29th June 2020. During this period, we managed to collect tweets that contain mentions to 3.15M users. We inspected the account visibility of each of those users using the Twitter user API, and found that around 4\%  of them were protected accounts at the time of the data collection. 

Our initial pool of protected users contain around 107K accounts. We applied a frequent check to get the account visibility of these protected accounts every 30 minutes for three months from 17th July until 17th October 2020\footnote{During this period, there were some instances where we could not reach some users because they were deactivated, suspended, or there were temporary problems with the Twitter API. There was only one instance where we could not reach any accounts because of a service disruption.}. Out of the 107K users that were protected on the first check, 64.5K (\char`~60\%) of them stayed protected during all three months. The remaining 42.5K (\char`~40\%) have switched their account visibility to public at least once during this period, including
7.6K (7.1\%) who changed to public and stayed that way.
Table~\ref{tab:changes} shows the number of switches users made in the three months we collected their account visibility. 

\begin{table*}[t]
	\centering
\begin{tabular}{c c c c c c c c } 
& 0 & 1 & 2 & 3 - 5 & 6 - 9 & 10+ & Total \\
\hline 
\# Users & 64,460& 7,630  & 9,195 & 8,599& 5,910 & 11,121 & 106,915\\ 
\end{tabular}
\caption{Number of switches users made in three months.}
\label{tab:changes}
\end{table*}

A user changing their tweet visibility settings frequently does not necessarily mean they stay protected/public equally. They can predominately stay protected or public and only switch for short time periods. Hence, we calculated how long our users stayed protected/public during the three months. For each user, we compared the number of 30 minutes blocks when they were protected to the number of blocks they were public. 14.6K (34.5\%) of the users who changed their account visibility at least once preferred to stay protected more than 95\% of the time. More than half of the switchers stayed protected 80\%+ of the time.

\subsection{Collection and Labeling of User Tweets}
We collected \change{the tweets of the switching users when their account was set to public. While it is possible to collect the last 3200 tweets of a user via Twitter API, we only collected the tweets posted during the account type check period (i.e.~between 17th July and 17th October 2020).} We managed to collect the tweets of 25.6K (60.4\%) users out of the 42.5K switching accounts in our collection.
Some of these users change their account visibility frequently and stay public for a short time. Hence, we could not get any tweets from the remaining 16.9K users.

\begin{figure}
    \centering
   \includegraphics[width=1\linewidth]{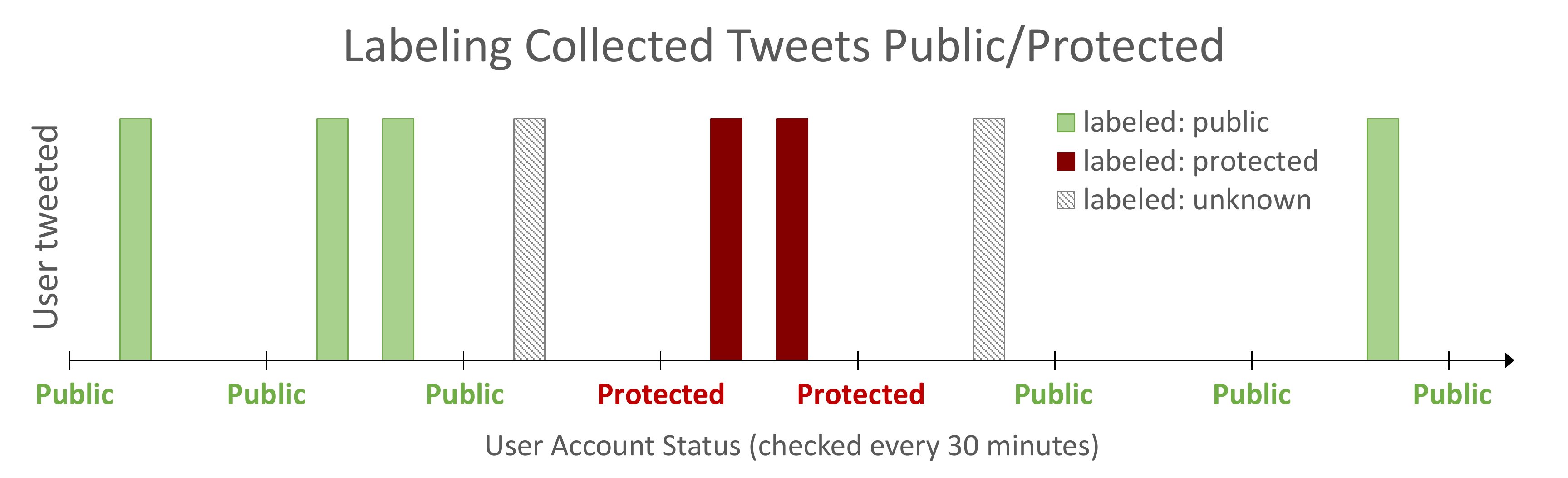}
   \Description[Tweet labeling process]{A bar chart with time-sequence showing the labeling process of tweets depending on account type of a user.}
\caption{\change{Example of the labeling process of the tweets of users in the dataset.}}
\label{fig:labeling}
\end{figure}

After collecting the tweets, and since we have the visibility status of the account every 30 minutes, we labeled them according to the collected account visibility of the user when the tweet was posted. If a tweet was sent at time $t$, we compared the user's account visibility collected just before and after $t$. If the account visibility settings were the same, we gave the tweet that label. In cases where the account setting visibility was different or we could not retrieve one of them, we discarded those tweets and did not include them in our analysis. \change{Figure~\ref{fig:labeling} illustrates an example of the labeling process of the tweets in a users' timeline}. For simplification purposes, we call tweets we labeled as public \emph{\change{\tweeto{}}} and the ones we labeled as protected \emph{\change{\tweetx{}}}.  \change{Table~\ref{tab:notations} gives the notations and definitions for the tweets we labeled}. Ultimately, we were able to label nearly 39.2M tweets (including retweets and replies) collected from the 25.6K switching accounts in the three months, where 26M (66.2\%) of them are \tweeto{} (public) while the remaining 13.2M (33.7\%) are \tweetx{} (protected). Nearly 9M \change{(23\%)} of these tweets were labeled as English by Twitter, and the remaining \change{77\% covered 64 }other languages \change{including Japanese, Portuguese, Indonesian, and Korean in that order}. 

\begin{table*}[t]
	\centering
\begin{tabular}{ c  p{12cm} } 
\change{Notation} & Definition \\
\hline 
\change{\tweeto{}} & Tweets labeled as \textbf{public} by our system, i.e.~user was public in our account type checks preceding and succeeding the tweet.\\
\change{\tweetx{}} & Tweets labeled as \textbf{protected} by our system, i.e.~user was protected in our account type checks preceding and succeeding the tweet.\\
\hline
\change{\usero{}} & Users who stayed public 90\%+ of the time we collected the account types.\\
\change{\userb{}} & Users who stayed protected between 10\% and 90\% of the time we collected the account types.\\
\change{\userx{}} & Users who stayed protected 90\%+ of the time we collected the account types.\\
\hline
\end{tabular}
\caption{\change{The notations and the definitions we use for users and labeled tweets of the Twitter data collection.}}
\label{tab:notations}
\end{table*}

Table~\ref{tab:duration} shows the user counts by percentages of the time the account was protected during the three months, as well as the number of total tweets we collected from these users. There is a positive relation between the time users stay protected and the number of tweets they send while they are protected. However, some users tend to tweet less when they were protected even if they stayed protected most of the time.

\begin{table*}[t]
	\centering
\begin{tabular}{l c c c c c } 
\% time as protected & [0-10]\% & (10-50]\% & (50-90)\% & 90+\% \\
\hline 
\# Users & 2.6K & 9.8K & 10.3K & 3K \\
\# \change{\tweeto{}} &5.3M & 14.3M& 6M & 375K \\
\# \change{\tweetx{}} & 214K & 3.4M & 7.1M& 2.5M \\
\# \tweeto{} $ + $ \tweetx{} & 5.5M & 17.7M & 13.1M& 2.9M\\

\end{tabular}
\caption{User counts by the percentage of time the account was protected during the three months of data collection. Also the total number of \tweeto{} and \tweetx{} collected for each group of users. \change{Only the users we could collect tweets from are reported.}}
\label{tab:duration}
\end{table*}

\subsection{Comparing tweets characteristics of users}
To understand the reasons why users switch their account visibility, we \change{divided them into three groups based on their privacy settings usage and analyzed them separately. These groups are \emph{\change{\userx{}}}, the }users who stay \emph{protected} 90\% of the time \change{or more}, \emph{\change{\usero{}}}, the users who stay \emph{public} 90\% of the time \change{or more}, \change{and \change{\userb{}}, the remaining users who have more \emph{balanced} duration of staying public or protected compared to \userx{} and \usero{}}. In this section, we compare \change{11 features} of the \tweetx{} and \tweeto{} tweets of \userx{}\change{, \userb{},} and \usero{}. \change{Table~\ref{tab:notations} shows the above notations and definitions for easy reference.}

As shown in Table~\ref{tab:duration}, out of the users we collected tweets from, 3K (11.7\%) of them were \userx{}, \change{20.1K (78.3\%) of them were \userb{},} and 2.6K (10\%) of them were \usero{}. 
\change{Users in our dataset }have the majority of their tweets with the same visibility state as the dominant setting.\\

\begin{table*}[t]
	\centering
	\small
\begin{tabular}{l l r c r c r } 
 & & $\overline{t_{x}}$ & 95\% CI & $\overline{t_{o}}$& 95\% CI & p-value \\
\hline 
&\change{Tweet Frequency} & \change{0.22} & \change{[0.21, 0.23]} & \change{0.59} & \change{[0.54, 0.63]} & \minorchange{***\textbf{\textless.001}}\\
&Mentions & 61.79 & \change{[60.99, 62.59]}& 64.11 & \change{[63.19 , 65.02]} & \minorchange{***\textbf{\textless.001}} \\
&Verified Ment. & 10.34 & \change{[9.63, 11.05]} & 15.30 & \change{[14.25, 16.35]} & \minorchange{***\textbf{\textless.001}}\\
&Non-follower Ment. & 52.28 & \change{[50.34, 54.23]} & 56.80 & \change{[54.66, 58.94]} & \minorchange{***\textbf{\textless.001}}\\
&Reply & 42.43 & \change{[41.48, 43.39]} & 43.29 & \change{[42.23, 44.35]} & \change{.022} \\
\userx{} &RT & 22.59 & \change{[21.68, 23.51]} & 22.02 & \change{[21.07, 22.96]} & .058 \\
&QT & 8.15 & \change{[7.80, 8.50]} & 8.39 & \change{[7.98, 8.81]} & .169 \\
&Urls & 7.99 & \change{[7.68, 8.30]} & 8.49 & \change{[8.07, 8.92]} & \change{.013}\\
&Hashtags & 3.70 & \change{[3.43, 3.97]} & 5.60 & \change{[5.15, 6.05]} & \minorchange{***\textbf{\textless.001}}  \\
&Media & 15.12 & \change{[14.61, 15.63]} & 15.07 & \change{[14.48, 15.65]} & .835 \\
&\change{English} & \change{17.06} & \change{[15.98, 18.14]} & \change{17.28} & \change{[16.17, 18.39]} & \change{.178} \\

\hline 
&\change{Tweet Frequency} & \change{0.28} & \change{[0.27, 0.28]} & \change{0.50} & \change{[0.48, 0.51]} & \minorchange{***\textbf{\textless.001}}\\

&\change{Mentions}  & \change{64.08} & \change{[63.77, 64.38]} & \change{65.73} & \change{[65.45, 66.02]} & \minorchange{***\textbf{\textless.001}} \\
&\change{Verified Ment.} & \change{11.00} & \change{[10.76, 11.25]} & \change{11.10} & \change{[10.86, 11.34]} & \change{.228} \\
&\change{Non-follower Ment.} & \change{57.16} & \change{[56.62, 57.71]} & \change{57.41} & \change{[56.89, 57.93]} & \change{.097} \\
&\change{Reply} & \change{41.31} & \change{[40.94, 41.68]} & \change{41.73} & \change{[41.38, 42.08]} & \minorchange{***\textbf{\textless.001}} \\
\change{\userb{}} &\change{RT} & \change{25.57} & \change{[25.20, 25.93]} & \change{25.19} & \change{[24.85, 25.53]} & \minorchange{***\textbf{\textless.001}}  \\
&\change{QT} & \change{8.53} & \change{[8.40, 8.66]} & \change{8.61} & \change{[8.48, 8.73]} & \change{.065} \\
&\change{Urls} & \change{8.36} & \change{[8.24, 8.49]} & \change{8.69} & \change{[8.56, 8.82]} & \minorchange{***\textbf{\textless.001}} \\
&\change{Hashtags} & \change{4.19} & \change{[4.07, 4.31]} & \change{5.16} & \change{[5.03, 5.29]} & \minorchange{***\textbf{\textless.001}}\\
&\change{Media} & \change{15.47} & \change{[15.27, 15.67]} & \change{15.80} & \change{[15.61, 15.98]} & \minorchange{***\textbf{\textless.001}} \\
&\change{English} & \change{19.55} & \change{[19.11, 19.99]} & \change{19.63} & \change{[19.19, 20.07]} & \change{.090} \\
\hline 
&\change{Tweet Frequency} & \change{0.47} & \change{[0.44, 0.50]} & \change{0.54} & \change{[0.51, 0.57]} & \minorchange{***\textbf{\textless.001}} \\

&Mentions & 66.43 & \change{[65.41, 67.44]} & 68.64 & \change{[67.86, 69.41]} & \minorchange{***\textbf{\textless.001}} \\
&Verified Ment. & 15.78 & \change{[14.69, 16.86]} & 13.36 & \change{[12.50, 14.23]} & \minorchange{***\textbf{\textless.001}}  \\
&Non-follower Ment. & 57.35 & \change{[55.89, 58.80]} & 55.20 & \change{[53.96, 56.45]} & \minorchange{***\textbf{\textless.001}} \\
&Reply & 40.35 & \change{[39.19, 41.52]} & 42.45 & \change{[41.47, 43.44]} & \minorchange{***\textbf{\textless.001}} \\
\usero{}&RT & 27.35 & \change{[26.20, 28.50]} & 26.78 & \change{[25.80, 27.77]} & .093 \\
&QT & 8.82 & \change{[8.33, 9.31]} & 8.94 & \change{[8.60, 9.28]} & .522 \\
&Urls & 8.82 & \change{[8.32, 9.32]} & 8.91 & \change{[8.57, 9.25]} & .673 \\
&Hashtags  & 4.82 & \change{[4.36, 5.28]} & 5.50 & \change{[5.14, 5.87]} & \change{*\textbf{.001}}\\
&Media & 16.99 & \change{[16.25, 17.72]} & 16.13 & \change{[15.62, 16.64]} & \change{.004}\\
&\change{English} & \change{23.05} & \change{[21.62, 24.48]} & \change{23.36} & \change{[21.98, 24.74]} & \change{.127} \\
\end{tabular}
\caption{Paired samples t-test comparing number of tweets of \userx{}, \change{\userb{},} and \usero{} when they are protected and public \change{(with Bonferroni correction *$p<(0.05/14)$, **$p<(0.01/14)$, ***$p<(0.001/14))$}.}
\label{tab:ttest}
\end{table*}

Here we compare the \change{tweeting frequencies, }language-independent tweet features, \change{and English tweeting behaviour of} \change{the three} user groups when they tweet under public or protected settings. 
Looking at tweet features can give us insights on why users might change their visibility settings back and forth. Hence, for each tweet we look at the following features: mentions (tweet has at least one mentioned username), verified mentions (mention of a verified user), non-follower mentions, reply (the tweet is a reply to another existing tweet), retweet (RT), quote tweet (QT), URLs, hashtags, media (e.g.~photo and video)\change{, and whether the tweet is in English using the language label given by the Twitter API}. We did not take the geo-location as a feature since only a small portion of the users had it enabled. \change{We collected the user profiles of the mentions to check whether they were verified. We also collected the followers of our users to check whether they mention non-followers in their tweets.\footnote{\change{The data collection for both of these features was done at a later time than the initial data collection. Considering Twitter networks are dynamic, the state of the network might have changed between these two collection times.}}} We extract these features from the tweets of \userx{}, \change{\userb{},} and \usero{}. We then calculate the percentage of prevalence in \tweetx{} and \tweeto{} for each of the features and compare them for every user. \change{We also divide the number of tweets a user sent while public/protected by the duration of staying public/protected so we can calculate the tweeting frequencies. We compare these features using paired sample t-test and apply Bonferroni correction post-hoc with p-values divided by 14.\footnote{\change{We made at most 14 comparisons for each user group but only report 11 of them in this paper.}}} 
For this analysis, 168\change{, 675,} and 227 users from \userx{}\change{, \userb{},} and \usero{} respectively were excluded due to all of their tweets being either public or protected.

Table~\ref{tab:ttest} shows the means of \change{values, 95\% confidence intervals,} and paired t-test p-values for each feature of the tweets for \userx{}\change{, \userb{},} and \usero{}. Users in \userx{} \change{tweet more when they are public. They also} mention Twitter users in their tweets more when they are public, albeit with a low effect size. The effect sizes are bigger for mentioning verified users and non-followers. Since a user's tweets can only be seen by followers, it is not surprising that they are switching to public to mention a verified account or someone who does not follow them.  Users in \userx{} also share more tweets with hashtags in them when they are public. They share media \change{and links} along with their tweets, \change{reply, }retweet, and quote tweet in similar rates. \change{The effect sizes for comparisons of tweeting frequency, verified user mentions, and non-followers mentions were small and the remaining comparisons all had very small effect sizes.}

\change{Similar to \userx{}, users in \userb{} tweet and mention other Twitter users in their tweets more while in the public setting. They also share media, links, and hashtags in their tweets and reply more when public. On the other hand, they retweet more when they are protected. Surprisingly, they mention verified users and non-followers in similar rates. They also tweet in English and quote tweet similarly when public and protected. Aside from the tweeting frequency, which had a small effect size, effect sizes were all very small for comparisons of \userb{}. }

\change{Users in \usero{} also tweet more when they are public. They }mention users, reply, \change{and share hashtags} more when they are public. However, surprisingly, they mention verified users and non-followers relatively more when they are protected. They \change{tweet in English, }retweet, quote tweet, \change{share media, and} links in similar rates. \change{The effect sizes for comparisons of \usero{} were all very small.}

\section{User Surveys}
The above data collection and analysis provides an interesting view on \emph{what} users did when they switch account visibility, but it does not provide an understanding of \emph{why} they are engaging in these actions. In this section we describe two surveys that we ran to better understand the reasons users change their settings and how common those reasons are. We start with an initial open-ended survey where we elicit reasons why people have switched in the past. We then combine the results of that survey with related work and our own observations during \change{the} data collection detailed in the previous section to create a list of common reasons people might change their visibility settings. We use the list in a second closed-ended survey to understand how common different switching reasons are. We give \change{the full} list and the sources of each reason in Appendix~\ref{app:reasons}.

\subsection{Identifying reasons to switch account visibility}

\begin{table*}[t]
\centering
\small
\begin{tabular}{p{4.5cm} c | p{4.5cm} c} 
Theme & \# users & Theme & \# users \\
\hline 
Sensitive/Political/Controversial & 19 & Giveaways* & 4\\
To prevent people I know from seeing  & 16 & Harassment & 3\\
Personal Information & 15 & Avoid suspension & 3 \\
To prevent non-followers from seeing & 10 & Viral Tweet & 3\\
Sense of privacy & 10 & Archival purposes & 2 \\
Broader audience* & 7 & NSFW & 1 \\
Interacting with non-followers* & 4 & & \\
\hline
\end{tabular}
\caption{Themes of free-text answers about reasons of switching. Most reasons are for switching from public to protected, while (*) indicates reason to switch from protected to public.}
\label{tab:free-text}
\end{table*}
The first survey \change{aims to identify} common reasons that lead users to switch between public and protected. \change{The survey is not intended to identify a comprehensive set of reasons for switching, and instead focuses on identifying common reasons which are then combined with information from literature (e.g.~\cite{marwick2014networked,rashidi2020s,veletsianos2018women}) to construct the options used in the next survey.}  

We recruited 100 Twitter users who have switched their tweet visibility at least once in the prior two months. The study was advertised as a ``Short survey about changing your tweets between public and protected on Twitter'' to Prolific Academic (PA)~\cite{prolific} users who are fluent English speakers \change{on March, 2021}. We conducted two pilots surveys with six participants each to adjust the wording of the questions as well as to get an accurate estimation of the time required. 100 participants took the final survey with an average completion time of 5.5 minutes and a compensation of £1.15. The study design was approved through our University's ethics process.

\emph{Survey Instrument:} 
The survey consisted of: informed consent, a screening question about if they had changed their Twitter account visibility (the ad clearly stated that this was required), four retrospective questions described below, demographics questions which included Twitter usage questions, and an optional free text comments box.  
For the retrospective questions we asked them to ``think back to a recent time when you have changed your Tweets from public to protected or from protected to public.'' The first question was free text and asked them what had motivated them to make the switch, followed by another free text question asking them what they hoped to achieve by making the switch. We also asked if the switch enabled them ``to achieve the effect I was trying for'' (Likert) and what direction the switch was in (to public, to protected, or multiple changes).

\emph{Analysis:} \change{Two researchers reviewed the responses and determined that the two free-text answers were often very related to each other, and that the answer for what they hoped to achieve often was an expansion of the motivation for the switch,
such that viewing them together provided a more comprehensive understanding of the reason. The answers were therefore analyzed together. We used an affinity diagram~\cite{HOLTZBLATT2017127} type approach to conduct a thematic analysis. The answers were placed in a shared spreadsheet, one researcher went through and sorted the answers into themes. They then met and the second researcher read through all the answers and adjusted the themes discussing with the first researcher as they went. The result was a set of themes that both researchers agreed on (Table~\ref{tab:free-text}}). 

\emph{Participants:} 
Participants were 23.6 years old on average with a max age of 39 years. 52 identified as male and 48 female. 65 described a situation where they switched to protected, 11 described switching to public, and 24 described an event that required multiple visibility changes.
When asked about the normal visibility of their account, 42\% indicated that they kept their account public most of the time, 27\% kept it mostly protected, while 31\% indicated that they switched more often. 62\% had also tweeted within the last week.

\subsubsection{Reasons to Switch:}

Participants most commonly switched to protected to talk about sensitive, political, or controversial topics freely. Reasons in this theme also overlapped with other themes, particularly around avoiding problematic interactions with strangers and proactively limiting the audience.
Another common reason was to prevent the people they know, such as friends and family, from seeing their tweets. Some of these users were uncomfortable because a person they knew found their account leading them to switch visibility. Other participants were uncomfortable with strangers or non-followers seeing their tweets. 
Participants also changed to protected temporarily to share personal information such as private Instagram links and then deleted the tweet before switching back to public. Harassment, having a tweet go viral, and sharing tweets that were not safe for work (NSFW) were some of the other reasons given. Uncommon but interesting reasons included wanting to archive an account and to avoid getting suspended due to complaints. 

Our participants primarily changed to public to reach a broader audience, interact with non-followers, as well as participating in giveaways where the account must be public to get selected for the prize. 
One participant cited all of these reasons when explaining why they changed to public. \textit{``I hoped to achieve what the platform could offer to its fullest, that included retweeting thing in order to comment on them, ``taking part'' in a trending hashtag or even taking part in giveaways.''}
Interestingly, one participant discussed going public for a while in order to gain more followers, then switching back to protected. Others said that it was challenging for friends and family to find them when they were protected, so they went public for a while to make themselves easier to find. We also asked if changing visibility enabled them to achieve the effect they were trying for and 89\% of the participants agreed that it did.

\subsection{Prevalence of Reasons}
The second survey looked at how common different reasons for switching visibility are. 
We used a combination of the results of the first survey, our observations from the Twitter data collection, and findings from related work to construct lists of reasons people switch to public and to protected. Then used the survey to find how many participants have switched due to these reasons. 

We advertised a study entitled ``5 min survey about changing your tweets between public and protected on Twitter'' on PA to fluent English speakers \change{on April, 2021}. The advertisement stated that we were looking for people who had changed their tweet visibility between public and protected two or more times in the last year to ensure respondents had experience with switching. 
We conducted a pilot survey with five participants to ensure accurate time estimates. 324 participants took the survey, requiring an average of 4 minutes to complete. They were compensated £0.75. Ethics approval was given by our University.

\emph{Survey Instrument:} 
Participants were first shown a consent page, followed by a screening to see if they had switched two or more times in the last year, followed by three multiple answer questions described next, questions about their Twitter experience, demographics, and an optional comment box.
The first two multiple answer questions asked users to indicate ``which of the following has previously lead you to change your tweet visibility settings to public?'' and also for protected. The third multiple answer question showed a list of protection activities, like deleting tweets when moving from protect to public, and asked users to select those they had used previously. We followed with a scenario-based question about a person who tweeted while protected and then switched to public and asked if their tweet was now public or protected.

\emph{Participants: }
Participants were an average of 25 years old ($\sigma=7.6$), with 46\% identifying as female, 53.4\% as male, and the remainder preferring not to say. 

Participants were asked what their normal public/protected balance was on their account. Ninety-nine (30.5\%) indicated their account was ``Always'' or ``Mostly'' \minorchange{public (Public} here forth), 28.4\% indicated their account was `Always'' or ``Mostly'' \minorchange{protected (Protected)}, and finally 41.0\% indicated that their visibility was ``Balanced'' or ``Somewhat'' public/protected \change{who are referred as Balanced users for the remainder of this section.} 

The majority (67.9\%) of participants accurately answered our scenario-based question about tweet visibility and said that the tweet sent while protected would now be publicly visible since the person switched to public. Fifty-three (16.4\%) thought that only users logged into Twitter could see it, but were aware that the tweet would be public \change{to all Twitter users}. But 15.7\% incorrectly thought that the tweet would remain protected.

We asked participants how often they had changed their visibility in the last three months.  
Sixty-two (19\%)  had not changed their tweet visibility settings in the last three months. One hundred and sixteen (35.8\%) had changed once, 27.5\% twice, 15.4\% three to five times, and 2.2\% changed six or more times. 

\subsubsection{Results:}

\begin{table*}[t]
	\centering
	\small
\begin{tabular}{p{9.5cm} c c c} 
Reasons to turn \textbf{public} & Public & Balanced & Protected\\
\hline 
To reach a broader audience and get more interaction with my tweets & 65.7\% & 50.4\% & 39.1\% \\
To mention/reply to a user who does not follow me & 49.5\% & 42.9\% & 41.3\%\\
To retweet other users & 49.5\% & 41.4\% & 31.5\%\\
To gain more followers & 43.4\%& 33.1\%&27.2\%\\
To quote tweet other users &33.3\% &27.1\%&21.7\%\\
To enter to get giveaways or freebies & 21.2\% & 21.8\% & 20.7\%\\
To share articles or links & 20.2\% & 20.3\%& 15.2\%\\
To share pictures & 23.2\%&15\%& 16.3\%\\
To associate a tweet with hashtags or trends publicly & 24.2\% & 12.8\% & 14.1\%\\
To boost the visibility, popularity, or ranking of a hashtag or topic & 25.3\% & 13.5\% & 8.7\%\\
To mention/reply to celebrities, famous people, or other VIPs & 17.2\% & 9.0\% & 8.7\%\\
To have a professional image & 7.1\% & 14.3\% & 6.5\%\\
To get customer service & 9.1\% & 9.8\% & 6.5\%\\
To boost the visibility of another user's tweet & 11.1\% & 6.8\%&8.7\%\\
To find potential employment & 0 & 5.3\% & 3.3\%\\
To sell things or receive donations & 4\% & 1.5\% & 0\\
Other & 1.0\% & 1.5\% & 0\\
I did not change my tweet visibility settings to public before & 2.0\% & 0.8\% & 7.6\%\\
\hline
\end{tabular}
\caption{Reasons to turn public.  Answers divided based on if their account is more often protected, public, or balanced. Percentages are out of the total number of mostly public, balanced, and mostly protected participants respectively.}
\label{tab:reasonsPublic}
\end{table*}

\begin{table*}[t]
	\centering
	\small
\begin{tabular}{p{9.5cm} c c c} 
Reasons to turn \textbf{protected} & Public & Balanced & Protected\\
\hline 
I wanted to prevent non-followers from seeing tweets with personal content & 50.5\% & 52.6\% & 63.0\%\\
People I know found my account and that made me uncomfortable  & 47.5\% & 51.9\% & 44.6\%\\
To get a sense of privacy & 35.4\% & 38.3\% & 46.7\%\\
To prevent people I know, such as friends and family, from seeing my tweets & 36.4\% & 38.3\% & 42.4\%\\
To prevent interactions from strangers & 30.3\% & 26.3\%& 38.0\%\\
To avoid harassment & 23.2\% & 27.1\% & 22.8\%\\
To talk about a sensitive, controversial, or political topics freely & 17.2\% & 15.8\% & 14.1\%\\
To take a temporary break from interactions with non-followers & 19.2\% & 14.3\% & 4.3\%\\ 
I did not want people to retweet me & 11.1\% & 8.3\% & 14.1\%\\
To archive the account without deleting it & 10.1\% & 9.8\% & 8.7\% \\
To tweet about someone without them being able to see the tweets & 9.1\% & 9.0\% &8.7\%\\
I did not want people to quote tweet me & 13.1\% & 6.0\% & 7.6\%\\
To share pictures &  5.1\% & 6.8\% & 10.9\%\\
My tweet unexpectedly went viral & 6.1\% & 2.3\% & 6.5\%\\
To share content that is not safe for work (NSFW) & 5.1\% & 4.5\% & 3.3\%\\
To quote tweet other users & 3.0\% & 4.5\% & 1.1\%\\
To prevent account suspension & 2.0\% & 2.3\% & 5.4\%\\
To share articles or links & 4.0\% & 0.8\% & 3.3\%\\
To retweet other users &  2.0\% & 3.0\% & 0\\
Other & 3.0\% & 1.5\% & 1.1\%\\
I did not change my tweet visibility settings to protected before & 1.0\% & 3.0\% & 0\\
\hline
\end{tabular}
\caption{Reasons to turn protected. Answers divided based on if their account is more often protected, public, or balanced. Percentages are out of the total number of mostly public, balanced, and mostly protected participants respectively.}
\label{tab:reasonsProtected}
\end{table*}

We asked participants to select all of the reasons that have caused them to change their visibility settings to public (Table~\ref{tab:reasonsPublic}) and protected (Table~\ref{tab:reasonsProtected}). On average users selected 3.3 ($\sigma=2.1$) protected reason options and 3.4 ($\sigma=2.0$) public reason options. 
The small number of selections and the range of selections shown in the table suggest that respondents were indeed reading the choices and only selecting those appropriate to them.

The most common reason to turn public for those that were mostly public or balanced was to ``reach a broader audience and get more interactions''. The finding makes sense given that interacting with tweets is a key functionality of Twitter. 
The most common set of reasons to turn public was to be able to interact with other accounts through mentioning, replying, retweeting, and quote tweeting. For those who were mostly protected, interactions like these seem to be the main driver to change visibility. 

The most common reason to turn protected for all groups was ``to prevent non-followers from seeing tweets with personal content''. In the first survey we saw several people switching to protected because someone they knew had found their account, and in this survey we similarly see that being a large reason to switch, even more so than preventing friends and family from seeing tweets, suggesting that the issue is around specific individuals more than just people they know. Preventing interaction from strangers and harassment were also common answers.  
``To get a sense of privacy'' was also a common answer, though interestingly not the most common. In the first survey, a common vague answer was that the user just wanted more privacy, which is why we added the answer option to the second survey despite its low specificity. 
Interestingly, sharing content like pictures, links, and not safe for work content were not common reasons to turn protected. Similarly, interacting with other users, which was a common reason to turn public, was not a common reason to turn protected.

\begin{table*}[t]
	\centering
	\small
\begin{tabular}{p{10cm} c c c} 
Control Visibility \& Interactions & Public & Balanced & Protected\\
\hline 
Delete some or all tweets when moving from protected to public & 49.5\% & 57.9\% & 52.2\% \\
Remove a follower without blocking them (soft blocking) & 46.5\% & 43.6\% & 54.3\%\\
Mute a follower so you can't see their interactions & 44.4\% & 39.8\% & 39.1\%\\
Block followers to prevent them from interacting even when your account is public & 34.3\% & 36.8\% & 33.7\%\\
Temporarily deactivate your account to prevent all interaction for a time & 17.2\% & 16.5\% & 10.9\%\\
Change to protected when not logged in or otherwise unable to respond to interactions & 12.1\% & 10.5\% &7.6\%\\
Have a clear list of engagement rules prominently shown or linked to that detail what is acceptable interaction or following behaviour & 4.0\% & 0.8\% & 1.1\%\\
Other & 0 & 0 & 1.1\%\\
I have not used any of the above & 6.1\% & 4.5\% & 3.3\%\\

\hline
\end{tabular}
\caption{Actions taken to control who can see and interact with tweets.  Answers divided based on if their account is more often protected, public, or balanced. Percentages are out of the total number of mostly public, balanced, and mostly protected participants respectively.}
\label{tab:controlStrategies}
\end{table*}

We also examined the combinations of answers users gave across \minorchange{these questions around turning public and protected}. The most frequent pair was ``To reach a broader audience and get more interaction with my tweets'' and ``I wanted to prevent non-followers from seeing tweets with personal content'' with 30.2\% of participants giving both those answers. 
Similarly, 18.5\% both indicated turning public to get more interactions while also indicating that they turned protected to prevent interactions from strangers. 
Another pair selected by 11.4\% participants was ``To enter to get giveaways or freebies'' and ``I wanted to prevent non-followers from seeing tweets with personal content''.

We also asked participants to select actions they have taken to control who can see and interact with their tweets (Table~\ref{tab:controlStrategies}).  
More than half indicated that they delete some or all of their tweets before changing their visibility to public. 
The control approaches of soft blocking, muting, and blocking were also used by many. Majority protected participants in particular preferred soft blocking, possibly because it removed followers without creating a notification to the impacted Twitter account and may be seen as less harsh. 

Interestingly, 23.1\% of participants indicated either temporarily deactivating their accounts to prevents interactions or changing to protected when they are not logged in and cannot respond to interactions. 
Twitter allows 30-day grace period for accounts that are deactivated, if a user re-activates their account in this time frame all content will be restored. 
We added these options based on prior research showing that some Facebook users delete their accounts when not online as a way of preventing interactions when they are unable to respond~\cite{marwick2014networked}. Based on these results, some Twitter users may be using a similar tactic.
Temporarily deactivating might also explain why we could not reach some accounts during the automated Twitter data collection.

\section{Discussion}

\subsection{\change{Summary of Findings}}

\change{We quantified privacy settings usage on Twitter to answer our RQ1 "How frequently do Twitter users change their tweet visibility settings?" After monitoring a set of 107K protected accounts on Twitter over three months, we 
found that 40\% of these accounts changed their visibility to public at least once over this period. Over a quarter of those who changed, did so 10 or more times. 
This shows an interesting behaviour of Twitter users when it comes to their use of tweet visibility settings, which motivated our RQ2 related to the posting strategies of these users when they are public vs protected.

Our analysis showed that users tweet less frequently when they are protected, even for those users who stayed mostly protected (\userx{}) during the data collection.
It was also interesting to find that all user groups mention others more when they are public. However, \usero{} mention verified users or non-followers more while protected, \userx{} mentions them more while public, while there is no change for \userb{}. This indicates that users change their usual setting to mention verified users and non-followers. 

We conducted two user surveys to answer the remaining research questions RQ3 and RQ4. We investigated the reasons behind the tweet visibility changes between public and protected to answer our RQ3. Over 40\% of our participants changed their tweet visibility settings to public so they can mention non-followers. Other interaction-based reasons such as retweeting, quote tweeting, and entering giveaways were popular reasons to turn public. On the other hand, our participants changed their tweet visibility to protected so they can limit the audience and unwanted interactions. Our participants were also wary of people they know, e.g.~family and friends, finding their accounts. Avoiding harassment and being able to talk freely about possibly controversial topics were among the most popular reasons to turn protected.

Our RQ4 concerns the strategies users utilize to manage their audience and interactions. The most popular strategy our participants employed was deleting tweets before moving the account from protected to public. Hidden preventions such as soft-blocking and muting were more popular compared to blocking, which can be noticed by the blocked person. Our participants also changed their account visibility to protected when they are unable to respond to interactions or deactivated their accounts so they can prevent all kinds of interactions including interactions from followers.

}

\subsection{\change{Implications}}
\label{sec:implications}

\change{Twitter's relatively simplistic privacy settings imply that accounts are either public, where anyone on the internet can see all their tweets, or protected, where only followers can see the tweets. An observation from this work is that for many users the visibility of their tweets is not a static setting but instead one that changes as their needs and circumstances change. For these users, it is inaccurate to think of them as simply public or protected as they are changing their settings to get a mix of the affordances granted by both those states. 

Current setting language also focuses around the protection of the tweets themselves rather than the person: 
``Unprotecting your Tweets will cause any previously protected Tweets to be made public''~\cite{twitter-public-protected}. 
We find that users made changes to settings to protect themselves (e.g.~harassment) or to create safe spaces (e.g.~discussing controversial topics), where the focus in both cases was around the users themselves and their ability to have conversations with other users in a private space rather than to necessarily protect the tweets. The overlap here is obviously large since the conversations they are having are happening via tweets so they have to protect the tweets in order to create a safe space to have these conversations. But the conceptual difference is quite important to understand as it contextualizes the different actions users are taking. 

Privacy expectations and notions of people are dynamic~\cite{altman1976conceptual} representing a continual effort to control their presentation of self in a social context~\cite{goffman1959presentation}. While Twitter's binary setting option is simple, which helps with understanding it, that simplicity also makes it more challenging to conduct fine grain boundary management. Particularly since users are not just moving themselves between a protected and public space, the way they might when they leave the house to go to a coffee shop, they are also bringing all their past tweets with them when they move between audience spaces, closer to holding an open house where anyone can enter the previously private space. The bringing of past events with them when they move between audience spaces makes proper boundary management more challenging since the whole tweet feed has to be curated to be appropriate for the new audience which can be cumbersome~\cite{vitak2014you,huang2020you,yilmaz2021perceptions}. Alternatively users may rely on self-censorship~\cite{marwick2011tweet,sleeper2013post} where they refrain from sharing sensitive information even when protected to limit the risks of it becoming public later. Approaches like these, or even lack of awareness, may be why 46\% of survey respondents indicated that they don't delete tweets when changing from protected to public. Another reason for not deleting past tweets might be the archival value of historical tweets and protecting meaningful self-representation~\cite{vitak2014you,huang2020you,yilmaz2021perceptions}.

Some of our users definitely curated their feeds when switching between protected and public. Deleting tweets was the most common audience control action selected in the survey, and in the tweet dataset we noticed fewer tweets when people were protected than when they were public which \emph{might} be an indication that users are deleting tweets before switching. Participants also mentioned the need to briefly protect a specific tweet or be able to briefly engage with a public account. For example, a participant mentioned being protected to share a private Instagram account with their followers to gain \textit{``a momentary bit of privacy before of deleting the tweet and making my account public again.''} Prior work has also shown that users delete their tweets for various reasons including preventing harassment, revealing too much information~\cite{sleeper2013read}, or simple typos~\cite{almuhimedi2013tweets}. In line with Mondal et al.~\cite{mondal2016forgetting}, we find that users utilize deletion to protect their privacy, control information flow, and manage audiences.

Users both want to ``achieve what the platform could offer to its fullest'' while also ``keep tweeting freely''. They are not just looking for a way to create a safe space, they also want to be able to interact with the wider community by doing things like replying to a public tweet or making parts of their tweet stream public.   Hence, we see the active management of tweet visibility in our dataset where we find that some users changed their settings more than 200 times in three months. The nature of the Twitter usage also seems to affect this behaviour. As a participant stated: \textit{``how much you change your public settings depends on which side of twitter you are. I have a fan account so I get more harassment than if I had a personal account and I change my setting more often''}.  

One solution our participants made use of was to have multiple accounts which had different settings, similar to Stutzman and Hartzog~\cite{stutzman2012boundary}. A participant stated: \textit{``in my twitter bubble it is very common to have two accounts, one public (and anonymous) and one private for the few close and trusted friends
made on the platform, talking about private matters and sharing images etc.''} Instagram users use ``finstas'' for similar purposes~\cite{xiao2020random}. Some users even appear to share their protected accounts on their bio and ask to be ``private moots (mutual follow)'' where they usually add only protected users to their network to minimize information leak, which can happen if they have conversations with public accounts~\cite{kekulluoglu2020analysing,kekulluoglu2022from}.}

\subsection{\change{Design Recommendations}}
\change{The largest design recommendation we have is that Twitter should think about tweet visibility in terms of conversations, time frames, or audiences rather than focusing on the account level only.}

\paragraph{\change{Visibility Pinning of Tweets:}}
\change{Provide users with the ability to ``pin'' a tweet as either public or protected so that it stays that way even when the account changes visibility. Doing so would allow users to more easily share a single tweet publicly or privately without needing to change their whole account visibility. It would also allow predominately protected users to share a public set of tweets that would allow them to attract new followers without opening their entire account up.}

\paragraph{\change{Paired Accounts:}}
\change{Directly support the practice of having both public and protected accounts. Users already use this approach to manage the visibility of their tweets and their audiences, but they are doing it ad hoc. Twitter could more directly support this practice by offering to create two linked accounts and making it easier to switch between them to post to appropriate audiences. Doing so would allow for better boundary management as well as assist users who are looking to gain followers by having some tweets public.} 

\paragraph{\change{Temporal Settings:}}
\change{Make some settings time based so that they encompass a time period or a conversation stream rather than just a single tweet. A user may change to protected to have a private conversation with their followers and then change back to public, but then they would need to go an delete all the tweets happening in that time frame. It would be helpful to users to have the ability to delete tweets that happened within a frame of time. It may also be helpful to have tweets that disappear after a set time frame or disappear when the account visibility changes. Posts with time-limits that disappear after a set time is already adopted by various social media platforms including Snapchat, Instagram, and WeChat. These features help users to cope with temporal context collapse to a degree and users appreciate these ephemeral posts~\cite{huang2020you}. However, the time-limit should be controlled by the users instead of the platform. According to Yilmaz et al.~\cite{yilmaz2021perceptions}, users found the automated deletion is beneficial only when the users are in control of the deletion date. Tweet deletion is also a common practice among users who do not necessarily switch their account visibility~\cite{mondal2016forgetting,almuhimedi2013tweets,sleeper2013read} so this feature may benefit them too.}

\paragraph{\change{Limiting Interactions:}}  \change{As there were participants who made their account public to interact, there were also participants who protected their accounts to limit interactions from non-followers. Some of our participants even deactivated their accounts temporarily to prevent all interactions for a while. There were also some participants who chose to limit interactions when they are unable to respond to them. Twitter recently introduced a feature for users to limit the replies to specific tweets. However, other interactions such as retweeting, quote tweeting, and liking are still possible for these tweets. Especially, quote tweets can be used as replies by other users when replies are limited, which also increases the reach of the original tweet more than the replies~\cite{garimella2016quote}. Users can stop all interaction by deactivating their accounts up to 30 days without deletion but deactivation will hide users' tweets, even from the user themselves. Enabling users to stop all interactions without changing their account visibility to protected or deactivating can help the participants adopting those strategies to limit/stop interactions. Additionally, prior work has shown that people see an archival value in their social media accounts~\cite{huang2020you,zhao2013many,vitak2014you}. This feature will benefit nearly 10\% of our participants who also wanted to archive their accounts without deleting it.}

\paragraph{\change{Removing Followers:}} 
\change{Allow users to remove followers directly rather than requiring them to soft-block them. Doing so would allow users to more easily curate their followers when they switch between public and protected. It would also be important to consider allowing them to do so silently without raising a notification to the other person since people prefer online sanctions that can create ``plausible deniability'' rather than visible ones, especially when dealing with followers who are strong-ties~\cite{rashidi2020s}.\footnote{\change{Twitter recently provided a feature for Twitter Web users to remove followers. However, this feature was introduced after we conducted our studies.}}}

\subsection{Limitations}

We curated our initial set of users by collecting mentioned protected accounts. Hence, our dataset is skewed towards users who stay protected most of the time. However, \change{we observed that the} users we collected tweets from \change{represent a wide range of percentage of time spent protected vs private as well as exhibiting a good distribution over that range.} Another limitation is the representation of mostly protected users in \tweetx{} and \tweeto{}. We \change{were only able to} collect tweets \change{of accounts when they are} public, \change{so in the case where a user} deletes their tweets before changing their tweets to public, \change{we would not be able to collect the deleted tweets}. Hence, for some users we might be getting only the \tweetx{} tweets that users are comfortable sharing publicly.

\change{Our survey participants also represent a younger population, who may have a different view on privacy and have different goals than people of other age groups. In the United States the median Twitter user age is 40 and 62\% of worldwide Twitter users are younger than 34~\cite{wojcik2019sizing, twitter-age}. Which means that our survey participants are indeed younger than a typical Twitter user but not greatly so.}
We recruited our survey participants from \change{the crowdsourcing platform Prolific Academic and crowdsource participants tend to be more privacy-conscious than the general population}~\cite{mturk-privacy}. This \change{population source may be the cause of} the higher percentage of mostly protected accounts in our participants (28\%) compared to their general percentage among Twitter users (5\%)~\cite{liu2014tweets}. Self reporting is \change{also} known to have biases, particularly around memory~\cite{moller13}, and our survey relies on answering questions about past events.

\change{The first survey was also intended to create an initial list of reasons that people switch between public and protected rather than create a comprehensive list. The number of participants surveyed, in our view, is high enough to identify common issues, but too few to really understand the scope of all possible reasons for switching. Also, as noted above, our participant sample is younger than an average Twitter user which also likely impacted the types of reasons they provided. We try and balance this issue by also including reasons found in prior work, but it is quite likely that if we were to more aggressively survey people from different age ranges, cultures, physical locations, or other demographics we would likely find a wider set of reasons for switching.}

\section{Conclusion}

In this study, we investigated Twitter users' privacy setting switching behaviour and the reasons behind the changes. To do so, we collected the account visibilities of 107K \change{initially protected} users for three months and found that nearly 40\% of them changed their settings at least once. We also collected the switchers' tweets to understand whether their sharing behaviour differs when they are protected vs public. We find that users utilize the privacy settings dynamically, sometimes changing as much as daily. They send tweets with mentions and hashtags more when they are public compared to protected. We coupled our Twitter data with two user surveys to get insight into the potential reasons behind the changes. We find that users change their accounts to protected to control their audience and interactions. On the other hand, users prefer to change their account visibility to public to use Twitter features freely, possibly at the expense of their privacy. We also suggest some design implications to protect the privacy of the users while enabling them to experience what the platform offers fully.

\begin{acks}
\minorchange{This work was supported in part by the EPSRC DTA award, funded by the UK Engineering and Physical Sciences Research Council and the University of Edinburgh. We thank everyone associated with the TULiPS Lab and SMASH Group at the University of Edinburgh for helpful discussions and feedback.}
\end{acks}

\balance
\bibliographystyle{ACM-Reference-Format}
\bibliography{references}


\clearpage
\onecolumn
\appendix
\section{Curation of Reasons for Switching Account Visibility}
\label{app:reasons}

\begin{table*}[h]
	\centering
	\small
\begin{tabular}{p{9.5cm} c c c} 
Reasons to turn \textbf{public} & Twitter Data & Free-Text Survey\\
\hline 
To reach a broader audience and get more interaction with my tweets & & +\\
To gain more followers & & + \\
To mention/reply to a user who does not follow me & + & +\\
To find potential employment &   & \\
To have a professional image & & +\\
To sell things or receive donations &  & \\
To enter to get giveaways or freebies &  & + \\
To retweet other users & + & +\\
To quote tweet other users & + & +\\
To associate a tweet with hashtags or trends publicly & + & + \\
To boost the visibility, popularity, or ranking of a hashtag or topic & & + \\
To boost the visibility of another user's tweet & & \\
To share articles or links & + & \\
To share pictures & + & +\\
To get customer service &  & \\
To mention/reply to celebrities, famous people, or other VIPs & + & + \\ \hline
\end{tabular}
\caption{Reasons to turn public - Options given in the survey and their sources}
\label{tab:reasonsPublicReference}
\end{table*}

\begin{table*}[h!]
	\centering
	\small
\begin{tabular}{p{9.5cm} c c c}
Reasons to turn \textbf{protected} & Twitter Data & Free-Text Survey\\
\hline 
People I know found my account and that made me uncomfortable  & & + \\
My tweet unexpectedly went viral & & +\\
I wanted to prevent non-followers from seeing tweets with personal content & & + \\
To prevent people I know, such as friends and family, from seeing my tweets & & + \\
To archive the account without deleting it & &+ \\
To avoid harassment & &+ \\
To tweet about someone without them being able to see the tweets & & +\\
To prevent account suspension &  & +\\
I did not want people to retweet me & + & \\
I did not want people to quote tweet me & + & \\
To talk about a sensitive, controversial, or political topics freely &  & +\\
To prevent interactions from strangers & & +\\
To share pictures & + &  + \\
To share content that is not safe for work (NSFW) &  & + \\
To retweet other users & + &  \\
To quote tweet other users & + & + \\
To get a sense of privacy &  & +\\
To share articles or links & + & \\
To take a temporary break from interactions with non-followers & & + \\ \hline
\end{tabular}
\caption{Reasons to turn protected - Options given in the survey and their sources}
\label{tab:reasonsProtectedReference}
\end{table*}

\end{document}